# Why Does Flow Director Cause Packet Reordering?

Wenji Wu, Phil DeMar, Matt Crawford

Fermilab, P.O. Box 500, Batavia, IL 60510

*Abstract* - Intel Ethernet Flow Director is an advanced network interface card (NIC) technology. It provides the benefits of parallel receive processing in multiprocessing environments and can automatically steer incoming network data to the same core on which its application process resides. However, our analysis and experiments show that Flow Director cannot guarantee in-order packet delivery in multiprocessing environments. Packet reordering causes various negative impacts. E.g., TCP performs poorly with severe packet reordering. In this paper, we use a simplified model to analyze why Flow Director can cause packet reordering. Our experiments verify our analysis.

## 1. Introduction

Computing is now shifting towards multiprocessing (e.g., CMP, SMP, and UNMA). The fundamental goal of multiprocessing is improved performance through the introduction of additional hardware threads, CPUs, or cores (all of which will be referred to as "cores" for simplicity). The emergence of multiprocessing has brought both opportunities and challenges for TCP/IP performance optimization in such environments. Modern network stacks can exploit parallel cores to allow either message-based parallelism or connection-based parallelism as a means of enhancing performance [1]. While existing OSes exploit parallelism by allowing multiple threads to carry out network operations concurrently in the kernel, supporting this parallelism carries significant costs, particularly in the context of contention for shared resources, software synchronization, and poor cache efficiencies [1][2]. Investigations regarding processor affinity [3][4][5] indicate that the coordinated affinity scheduling of protocol processing and network applications on the same target cores can significantly reduce contention for shared resources, minimize software synchronization overheads, and enhance cache efficiency.

Coordinated affinity scheduling of protocol processing and network applications on the same target cores has the following goals: (1) *Interrupt affinity:* Network interrupts of the same type should be directed to a single core. Redistributing network interrupts in either a random or round-robin fashion to different cores has undesirable side effects [4]. (2) *Flow affinity*: Packets belonging to a specific flow should be processed by the same core. Flow affinity is especially important for TCP. TCP is a connection-oriented protocol, and it has a large and frequently accessed state that must be shared and protected when packets from the same connection are processed. Ensuring that all packets in a TCP flow are processed by a single core reduces contention for shared resources, minimizes software synchronization, and enhances cache efficiency. (3) *Network data affinity*: Incoming network data should be steered to the same core on which its application process resides. This is becoming more important with the advent of Direct Cache Access (DCA) [6]. Network data affinity maximizes cache efficiency and reduces core-to-core synchronization. In a multicore system, the function of network data steering is executed by directing the corresponding network interrupts to a specific core (or cores).

Receive Side Scaling (RSS) [7] is a NIC technology. It supports multiple receive queues and integrates a hashing function in the NIC. The NIC computes a hash value for each incoming packet. Based on hash values, NIC assigns packets of the same data flow to a single queue and evenly distributes traffic flows across queues. With Message Signal Interrupt (MSI/MSI-X) [8] support, each receive queue is assigned a dedicated interrupt and RSS steers interrupts on a per-queue basis. RSS provides the benefits of parallel receive processing in multiprocessing environments. Operating systems like Windows, Solaris, Linux, and FreeBSD now support interrupt affinity. When an RSS receive queue (or interrupt) is tied to a specific core, packets from the same flow are steered to that core (Flow pinning [9]). This ensures flow affinity on most OSes, with Linux being the major exception. However, RSS has a limitation: it cannot steer incoming network data to the same core where its application process resides. The reason is simple: the existing RSS-enabled NICs do not maintain the relationship in the NIC:

Traffic Flows → Network applications → Cores

Since network applications run on cores, the most critical relationship is simply:

Traffic Flows → Cores (Applications)

Unfortunately, RSS does not support such capability. This is symptomatic of a broader disconnect between existing software architecture and multicore hardware. With OSes like Windows and Linux, if an application is running on one core, while RSS has scheduled received traffic to be processed on a different core, poor cache efficiency and significant core-to-core synchronization overheads will result. The overall system efficiency may be severely degraded. To remedy the RSS limitation, the Intel Ethernet Flow Director technology [10] has been introduced. The basic idea is simple: Flow Director maintains the relationship "Traffic Flows → Cores (Applications)" in the NIC. OSes are correspondingly enhanced to support such capability. Flow Director not only provides the benefits of parallel receive processing in multiprocessing environments, it also can automatically steer packets of a specific data flow to the same core, where they will be protocol-processed and finally consumed by the application. However, our analysis and experiments show that Flow Director cannot guarantee in-order packet delivery in multiprocessing environments. TCP performance suffers in the event of severe packet reordering. In this paper, we use a simplified model to analyze why Flow Director can cause packet reordering. Our experiments verify our analysis.

## 2. Why does Flow Director Cause Packet Reordering?

Intel Ethernet Flow Director [10] is a NIC technology. It supports multiple receive queues in the NIC, up to the number of cores in the system. With MSI/MSI-X and Flow-Pinning support, each receive queue has a dedicated interrupt and is tied to a specific core; each core in the system is assigned a specific receive queue. The NIC device driver allocates and maintains a ring buffer in system memory for each receive queue. For packet reception, a ring buffer must be initialized and pre-allocated with empty packet buffers that have been memory-mapped into address space accessible by the NIC over the system I/O bus. The ring buffer size is device and driver-dependent. For Flow-Director-steering traffic, Flow Director maintains a "Traffic Flow → Core" table with a single entry per flow. Each entry tracks the receive queue (core) to which a flow should be assigned. Entries within the "Traffic Flow → Core" table are updated by outgoing packets. To support Flow Director, OS must be multiple TX queue capable [11]. Each core in the system is assigned a specific transmit queue. Outgoing traffic generated on a specific core is transmitted via its corresponding transmit queue. For an outgoing transport-layer packet, the OS records a processing core ID and pass it to the NIC to update the corresponding entry in the table. Flow Director makes use of the 5-tuple *{src_addr, dst_addr, protocol, src_port, dst_port}* in the receive direction to specify a flow. Therefore, for an outgoing packet with the header *{(src_addr: x), (dst_addr: y), (protocol: z), (src_port: p), (dst_port: q)}*, its corresponding flow entry in the table is identified as *{(src_addr: y), (dst_addr: x), (protocol: z), (src_port: q), (dst_port: p)}*.

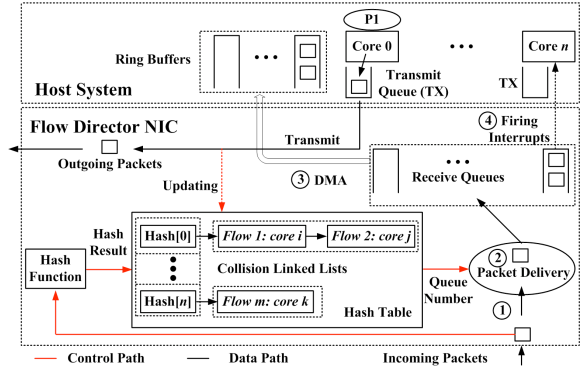

**Fig. 1 Flow Director Mechanism**

Fig. 1 illustrates packet receive-processing for transport-layer packets with Flow Director. (1) When incoming packets arrive, the hash function is applied to the header to produce a hash result. Based on the hash result, the NIC identifies the core and hence, the associated receive queue. (2) The NIC assigns the incoming packets to the corresponding receive queues. (3) The NIC deposits via direct memory access (DMA) the received packets into the corresponding ring buffers in system memory. (4) The NIC sends interrupts to the cores associated with the non-empty queues. Subsequently, the cores respond to the network interrupts and process the received packets up through the network stack from the corresponding ring buffers one by one. As for non-Flow-Director-steering traffic, please refer to [10] for more details.

Flow Director not only provides the benefits of parallel receive processing in multiprocessing environments, it also can automatically steer packets of a data flow to the same core, where they will be protocol-processed and finally consumed by the application. However, our analysis shows that Flow Director cannot guarantee in-order packet delivery in multiprocessing environments. TCP performs poorly with severe packet reordering [12]. In the following section, we use a simplified model to analyze why Flow Director cannot guarantee in-order packet delivery.

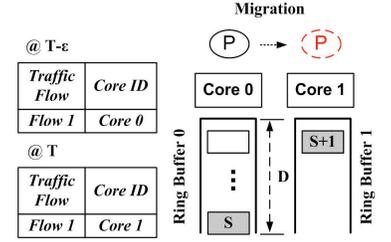

**Fig. 2 A Simplified Model for Packet Reordering Analysis**

As shown in Fig. 2, at time $T-\varepsilon$, *Flow 1*'s flow entry maps to Core 0 in the "Traffic Flow → Core" table. At this instant, packet *S* of *Flow 1* arrives; based on the "Traffic Flow → Core" table, it is assigned to Core 0. At time $T$, due to process migration, *Flow 1*'s flow entry is updated and maps to Core 1. At $T+\varepsilon$, Packet *S+1* of *Flow 1* arrives and is assigned to the new core, namely Core 1. As described above, after assigning received packets to the corresponding receive queues, NIC copies them into system memory via DMA, and fires network interrupts, if necessary. When a core responds to a network interrupt, it processes received packets up through the network stack from the corresponding ring buffer one by one. In our case, Core 0 processes packet S up through the network stack from Ring Buffer 0, and Core 1 services packet S+1 from Ring Buffer 1. Let $T_{service}(S)$ and $T_{service}(S+1)$ be the times at which the network stack starts to service packets S and S+1, respectively. If $T_{service}(S) > T_{service}(S+1)$, the network stack would receive packet S+1 earlier than packet S, resulting in packet reordering. Let D be the ring buffer size and let the network stack's packet service rate be $R_{service}$ (packets per second). Assume there are $n$ packets ahead of S in Ring Buffer 0 and $m$ packets ahead of S+1 in Ring Buffer 1. Then:

$$T_{service}(S) = T - \varepsilon + n/R_{service} \qquad (1)$$
$$T_{service}(S+1) = T + \varepsilon + m/R_{service} \qquad (2)$$

If $\varepsilon$ is small and $n > m$, the condition of $T_{service}(S) > T_{service}(S+1)$ would easily hold and lead to packet reordering. Since the ring buffer size is $D$, the worst case is $n = D-1$ and $m = 0$:

$$T_{service}(S) = T - \varepsilon + (D-1)/R_{service} \qquad (3)$$
$$T_{service}(S+1) = T + \varepsilon \qquad (4)$$

The ring buffer size D is a design parameter for the NIC and driver. For example, the Myricom 10Gb NIC is 512, and Intel's 1Gb NIC is 256.

In a multicore system, a general-purpose OS scheduler tries to use all core resources in parallel as much as possible, distributing and adjusting the load among the cores. Process migration across cores occurs frequently. Flow Director can easily cause packet reordering in these conditions.

To validate our analysis, we ran data transmission experiments over an isolated network. A sender was directly connected to a receiver via a physical 10Gbps link. The sender and receiver's detailed features were:

**Sender:** Dell R-805. CPU: two Quad Core AMD Opteron 2346HE, 1.8GHz. NIC: Myricom 10Gbps Ethernet NIC. OS: Linux 2.6.28.

**Receiver:** SuperMicro Server. CPU: two Intel Xeon CPUs, 2.66 GHz, Family 6, Model 15. NIC: Intel X520 Server Adapter with Flow Director enabled (configured with suggested default parameters [11]: FdirMode=1, AtrSampleRate=20), 10Gbps. OS: Linux 2.6.34, Multiple TX Queue Capable.

In our experiments, *iperf* [13] is used to send *n* parallel TCP streams from sender to receiver for 100 seconds. We ran "iperf –s" in the receiver. Linux was configured to run in *multicore peak performance* mode [14]. As a consequence, the scheduler tries to use all core resources in parallel as much as possible, distributing the load equally among the cores. Iperf is a multi-threaded network application. With multiple parallel TCP data streams, a dedicated child thread is spawned and assigned to handle each stream. As a result, iperf threads may migrate across cores. The receiver was instrumented to record out-of-order packets, and we calculated relevant packet reordering ratios. The experiment results, with a 95% confidence interval, are shown in Table 1.

| n | Reordering Ratio |
|---|---|
| 40 | 0.498% ± 0.067% |
| 100 | 0.705% ± 0.042% |
| 200 | 0.897% ± 0.038% |
| 500 | 0.635% ± 0.154% |
| 1000 | 0.409% ± 0.009% |
| 2000 | 0.129% ± 0.003% |

**Table 1 Experiment Results**

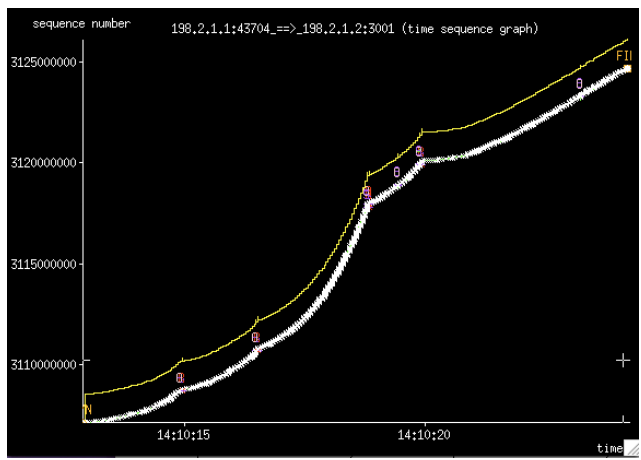

**Fig.3 Packet Trace Analysis (@*n=200*)**

The degree of packet reordering is significant. At *n=200*, packet reordering ratio reaches as high as 0.897%. The experiment results validated our analysis. When the scheduler tries to use all core resources in parallel as much as possible, distributing the load equally among the cores, it will lead to frequent process migration. As our analysis suggested, the Flow Director mechanism would cause packet reordering when process migration occurs. In addition, we ran tcpdump to record a single stream's packet trace at the receiver @ *n=200*. The packet trace analysis in Fig. 3 (using tcptrace and xplot [15]) clearly shows the occurrence of duplicate ACKs, SACKs, and data retransmissions due to packet reordering.

We then ran "taskset 0x01 iperf –s" in the receiver to pin iperf to core 0 and repeated the above experiments. No packet reordering was discovered. This is because when iperf is pinned to a specific core, its child threads are also pinned to that core. There will be no process migration in this case. In these conditions, Flow Director does not cause packet reordering.

## 3. Conclusion

In this paper, we use a simplified model to analyze why Flow Director can cause packet reordering in multiprocessing environments. Our experiments validate our analysis. The contributions of this paper are twofold. First, we show Intel Ethernet Flow Director cannot guarantee in-order packet delivery in multiprocessing environments. Second, we develop a simplified model to analyze why Flow Director causes packet reordering.

## REFERENCE

[1] P. Willmann et al., "An Evaluation of Network Stack Parallelization Strategies in Modern Operating Systems," USENIX ATC, 2006.

[2] J. Hurwitz et al., "End-to-end performance of 10-gigabit etherent on commodity systems," IEEE Micro, Vol. 24, No. 1, 2004, pp. 10-22.

[3] J. Salehi et al., "The effectiveness of affinity-based scheduling in multiprocessor networking," IEEE/ACM Transactions on Networking, Volume 4, Issue 4, 1996 Page(s): 516-530.

[4] A. Foong et al., "An in-depth analysis of the impact of processor affinity on network performance," In *Proc. IEEE International* Conference on Networks, 2004.

[5] J. Hye-Churn et al., "MiAMI: Multi-Core Aware Processor Affinity for TCP/IP over Multiple Network Interfaces," In Proc. IEEE Symposium on High Performance Interconnects, 2009.

[6] R. Huggahalli et al., "Direct Cache Access for High Bandwidth Network I/O," In Proc. 32nd Annual International Symposium on Computer Architecture, 2005.

[7] www.microsoft.com, Receive-Side Scaling Enhancements in Windows Server, 2008.

[8] PCI Express System Architecture: PC System Architecture Series, Addison-Wesley Professional, 2003, ISBN-10: 0321156307.

[9] www.intel.com, Supra-linear Packet Processing Performance with Intel Multi-core, 2006.

[10] Intel 82599 10GbE Controller Datasheet, 2009.

[11] www.intel.com, IXGBE device driver README.

[12] W. Wu et al., "Sorting reordered packets with interrupt coalescing," computer network, Volume 53, Issue 15, 2009, pages: 2646-2662.

[13] http://dast.nlanr.net/Projects/Iperf/

[14] www.kernel.org

[15] www.tcptrace.org